\pdfoutput=1

\documentclass[11pt]{article}
\RequirePackage[OT1]{fontenc}
\RequirePackage{amsthm,amsmath}
\RequirePackage[numbers]{natbib}
\RequirePackage[colorlinks,citecolor=blue,urlcolor=blue]{hyperref}

\usepackage[utf8]{inputenc} 
\usepackage[T1]{fontenc}    
\usepackage[colorlinks]{hyperref} 
\usepackage{url}            
\usepackage{booktabs}       
\usepackage{amsfonts}       
\usepackage{amsmath}
\usepackage{nicefrac}       
\usepackage{microtype}      
\usepackage{color}
\usepackage{bbm}
\usepackage{amssymb, enumerate}
\usepackage{makecell}
\usepackage{wrapfig}
\usepackage{extarrows}
\usepackage{arydshln}

\usepackage{caption}
\usepackage{multirow}

\usepackage{amsmath,amsfonts,amsthm, amssymb, enumerate}
\usepackage{lipsum}
\usepackage{caption}
\usepackage{amssymb,graphicx,subfigure}
\usepackage[bottom]{footmisc}
\usepackage[dvipsnames]{xcolor}
\usepackage{array,multirow}
\usepackage{enumitem}

\usepackage{microtype}
\usepackage{graphicx}
\usepackage{subfigure}
\usepackage{booktabs} 
 
\usepackage{graphicx}
\usepackage{amsmath}
\usepackage{amssymb}
\usepackage{booktabs}  
\usepackage{times}
\usepackage{epsfig}
\usepackage{graphicx}
\usepackage{amsfonts, amsthm, enumerate} 
\usepackage[bottom]{footmisc} 
\usepackage{enumitem}
\usepackage{setspace}
\usepackage{wrapfig}
\usepackage{extarrows}
\usepackage{makecell}
\usepackage{bbm}  
\usepackage{textcomp}
\usepackage{xcolor} 
\usepackage{caption}

\usepackage[font=small,labelfont=bf]{caption}

\usepackage{algorithm,algpseudocode}
\algnewcommand\TR{\item[{\textbf{Training phase}}]}
\algnewcommand\TE{\item[{\textbf{Test phase}}]}
\algnewcommand\Input{\item[{{Input:}}]}
\algnewcommand\Output{\item[{{Output:}}]}
\algnewcommand\Initialize{\item[{{Initialize:}}]}
\algnewcommand{\return}[1]{
	\State \textbf{return:}
	\Statex \hspace*{\algorithmicindent}\parbox[t]{.8\linewidth}{\raggedright #1}
}


\usepackage{setspace}
\begin{document}
	\title{AI-based computer-aided diagnostic system of chest digital tomography synthesis: Demonstrating comparative advantage with X-ray-based AI systems}
	\date{}
	\author{
		Kyung-Su Kim$^{1,2}$\thanks{Equal contribution}\,\,\thanks{Corresponding author: Kyung-Su Kim (kskim.doc@gmail.com) and Myung Jin Chung (mj1.chung@samsung.com)}, Ju Hwan Lee$^{3}$\footnotemark[1], Seong Je Oh$^{3}$\footnotemark[1], Myung Jin Chung$^{1,2,4}$\footnotemark[2]\\ 
		{\small $^{1}$Medical AI Research Center, Research Institute for Future Medicine, Samsung Medical Center, Seoul, Korea}\\
		{\small $^{2}$Department of Data Convergence and Future Medicine, Sungkyunkwan University School of Medicine, Seoul, Korea}\\
		{\small $^{3}$Department of Health Sciences and Technology, SAIHST, Sungkyunkwan University, Seoul, Korea}\\
		{\small $^{4}$Department of Radiology, Samsung Medical Center, Sungkyunkwan University School of Medicine, Seoul, Korea}
	}

	\maketitle 
	
\begin{abstract}
\textbf{Background} $\,\,$
Compared with chest X-ray (CXR) imaging, which is a single image projected from the front of the patient, chest digital tomosynthesis (CDTS) imaging can be more advantageous for lung lesion detection because it acquires multiple images projected from multiple angles of the patient. Various clinical comparative analysis and verification studies have been reported to demonstrate this, but there are no artificial intelligence (AI)-based comparative analysis studies. Existing AI-based computer-aided detection (CAD) systems for lung lesion diagnosis have been developed mainly based on CXR images; however, CAD-based on CDTS, which uses multi-angle images of patients in various directions, has not been proposed verified for its usefulness compared to CXR-based counterparts. \\
\textbf{Background and Objective} $\,\,$
This study develops and tests a CDTS-based AI CAD system to detect lung lesions to demonstrate performance improvements compared to CXR-based AI CAD. \\
\textbf{Methods} $\,\,$
We used multiple (e.g., five) projection images as input for the CDTS-based AI model and a single-projection image as input for CXR-based AI model to compare and evaluate the performance between models. Multiple/single projection input images were obtained by virtual projection on the {three-dimensional} (3D) stack of computed tomography (CT) slices of each patient's lungs from which the bed area was removed. These multiple images result from shooting from the front and left and right 30/60 °. The projected image captured from the front was used as the input for the CXR-based AI model. The CDTS-based AI model used all five projected images. The proposed CDTS-based AI model consisted of five AI models that received images in each of the five directions, and obtained the final prediction result through an ensemble of five models. Each model used WideResNet-50. To train and evaluate CXR- and CDTS-based AI models, 500 healthy data, 206 tuberculosis data, and 242 pneumonia data were used, and {three} three-fold cross-validation was applied. \\
\textbf{Results} $\,\,$
The proposed CDTS-based AI CAD system yielded sensitivities of 0.782 and 0.785 and accuracies of 0.895 and 0.837 for the (binary classification) performance of detecting tuberculosis and pneumonia, respectively, against normal subjects. These results show higher performance than the sensitivity of 0.728 and 0.698 and accuracies of 0.874 and 0.826 for detecting tuberculosis and pneumonia through the CXR-based AI CAD, which only uses a single projection image in the frontal direction. We found that CDTS-based AI CAD improved the sensitivity of tuberculosis and pneumonia by 5.4$\%$ and 8.7$\%$ respectively, compared to CXR-based {AI} CAD without loss of accuracy. \\
\textbf{Conclusions} $\,\,$
This study comparatively proves that CDTS-based AI CAD technology can improve performance more than CXR. These results suggest that we can enhance the clinical application of CDTS. Our code is available at \url{https://github.com/kskim-phd/CDTS-CAD-P}. 
\end{abstract}

\section{Introduction}
\subsection{Importance of CXR and its limitations}
\,\,\,\,\,\,\,\, CXR images the human chest area using only X-rays, without inserting any substances (e.g., contrast agents and instruments) into the body. As it has less inexpensive, is more accessible, and can be shot in a shorter time than other imaging modalities (e.g., chest CT) \cite{self2013high,woodring1986update}, the CXR is the most widely used first-line/non-invasive lung lesion diagnostic method \cite{self2013high,chandra2021coronavirus}. Specifically, as the American College of Radiology reported that the approximate effective radiation doses of CT and X-ray of the chest were 6.1 and 0.1 msv respectively \cite{safety2012radiation}, the radiation detection amount of CXR is 6msv below that of chest CT so rapid examination is possible with a small radiation dose through CXR \cite{zhu2004low}. Despite this low dose, CXR can effectively capture important information (e.g., lungs, heart, chest wall, and clavicle) inside the chest, making it widely used as a first-time examination device \cite{palmer1979pulmonary}.

However, CXR is less sensitive to disease diagnosis than CT \cite{langer2016sensitivity,choi2020missed,quaia2014diagnostic}; therefore, there are still many limitations in using it for precise diagnosis other than for the first-line test. A CXR is a two-dimensional {(2D)} image of a patient's {3D} lung area projected in a simple frontal direction. This characteristic makes lesions superimposed on other structures (e.g., the diaphragm, heart, and ribs), which lowers the sensitivity of CXR to specific diseases \cite{choi2020missed,tang2021disentangled,whiting2015computed,bachman1978effects}.

\subsection{Importance of DTS as an alternative to CXR}
\,\,\,\,\,\,\,\, Digital tomosynthesis (DTS) reconstructs multi-projection images obtained by moving an X-ray tube within a limited angle range into a 3D image, similar to CT \cite{ren2008novel}, used for breast image restoration in the name of digital mammography. CDTS refers to DTS targeting the lung area, and it can compensate for the difficulty of lesion diagnosis and the inability to determine the depth of the lesion on CXR. Unlike CXR, CDTS can separate overlapping anatomical structures into subsequent slices, such as CT, visualizing additional abnormalities that cannot be expressed by CXR \cite{yamada2013tomosynthesis}.   Based on these features, several clinical studies have reported that CDTS can improve lung lesion sensitivity compared to conventional CXR \cite{cant2017can,langer2016sensitivity,johnsson2012pulmonary,kim2010pulmonary,chawla2009design}. Specifically, Kim et al.\cite{kim2010pulmonary} clinically verified that the sensitivity of DTS was superior to CXR in pulmonary mycobacterial disease and reported that out of 141 pulmonary cavities detected on CT of 100 patients with the pulmonary mycobacterial disease, an average of 27 (19\%) in CXR and an average of 108 (77\%) in DTS. Unlike other lesions, lung lesions are small and often inconspicuous because of the anatomical structures around them \cite{larici2017lung,stitik1978radiographic}, so CDTS can have an advantage over CXR, especially in precisely diagnosing lung lesions. That is, CDTS improved detection sensitivity than CXR while offering a lower radiation dose, and lower cost compared to CT \cite{dobbins2003digital}, thereby making it suitable for the first-time equipment for more precise diagnosis of lung lesions as an effective alternative to CXR.

\subsection{Contribution}
\,\,\,\,\,\,\,\, As verified by mass or microcalcification detection in mammograms \cite{lotter2017multi,chan2005computer,samala2016deep}, the superiority of DTS-based AI CAD has recently been in the spotlight; however, the superiority of CDTS-based AI CAD technology still has not been appropriately compared and verified. Although there was a case where CDTS-based AI CAD was developed \cite{Chauvie2020ArtificialIA}, this technology only compared performance with non-AI-based CAD technologies (e.g., logistic regression and random forest), not with CXR AI-based CAD. CXR AI-based CAD technology is the most popular technology recently \cite{shorfuzzaman2021artificial, zhang2021pneumonia}, and CXR is closely related to CDTS in terms of being based on a radiation-based projection image; therefore, many comparative verifications have been made clinically, but fair comparative verification based on AI CAD has not yet been performed. This study developed an AI CAD for CDTS and, for the first time, evaluated its diagnostic efficacy in comparison with CXR. The contributions and characteristics of this study are: 

\begin{itemize}[noitemsep,topsep=0pt,leftmargin=3.5mm]
    \item For a fair comparison, we developed an AI CAD for CDTS based on its multi-projection images before restoration. We developed an AI CAD model applicable to any number of multi-projection images and then compared and evaluated whether this model using all multi-projection images (i.e., CDTS-based AI CAD) outperforms that using only a frontal image (i.e., CXR-based AI CAD). Our study has a structure that only transforms input data for the same model, enabling fair AI CAD performance comparison between CDTS and CXR.
    \item As DTS reconstructs multi-projection images obtained within a limited angle into 3D images, significant information loss occurs during the under-sampled inversion process \cite{sidky2009enhanced,velikina2007limited,hu2008image}. Unlike existing AI technologies \cite{Chauvie2020ArtificialIA,lotter2017multi,chan2005computer,samala2016deep} which are based on images after restoration, the proposed CDTS-based AI CAD minimizes the information loss as it is an AI CAD based on multi-projection images, which are images before restoration.
    \item Our CDTS-based AI CAD was developed based on the $N/A$ method. The proposed $N/A$ method assumes an AI network that inputs each of the images projected from $N$ varying directions as an input. Then the $N/A$ method refers to an algorithm that determines the final positive result when $A$ or more of the $N$ total diagnostic results, which are individual outputs of these $N$ networks, are judged to be positive. Unlike the existing ensemble method such as majority, vote \cite{ferreira2020multi}, the proposed $N/A$ method is clinically more valuable as it controls the sensitivity of disease determination (with adjustment of the value of $A$).
    \item We also developed a novel disease visualization technique suitable for the proposed CDTS-based AI CAD, based on a lung segmentation mask-based class activation map (CAM) method that visualizes disease regions more reliably in each projection image. This is achieved by developing a segmentation network that receives each projection image as an input and is learned to provide the lung region mask of each image as an output. We allow the CAM of $N$ individual diagnostic models to activate only the corresponding predicted lung regions. Our visualization technique enables the network to express discriminative areas of lung disease more intensively than existing CAM through this process.
    \item We performed an anomaly detection experiment for tuberculosis and pneumonia and observed that CDTS-based AI CAD (Sensitivity 0.782 and 0.785 / Accuracy 0.895 and 0.837) significantly outperforms CXR-based AI CAD (Sensitivity 0.728 and 0.698 / Accuracy 0.874 and 0.826) respectively, especially improving lesion detection sensitivity by at least 5\% than CXR-based one without loss of accuracy. Therefore applying anomaly detection to tuberculosis and pneumonia, the lesion sensitivities of the frontal-only CXR AI CAD were 0.728 and 0.698, whereas the lesion sensitivity of our proposed CDTS AI CAD was 0.782 and 0.785, achieving higher lesion sensitivity. These results showed that CDTS AI CAD technology could also be superior and valuable to CXR AI CAD technology in detecting lung abnormalities, like the results of existing clinical observational studies.
\end{itemize}

\begin{figure}[ht]
\centering
\includegraphics[width=\linewidth, height=9cm]{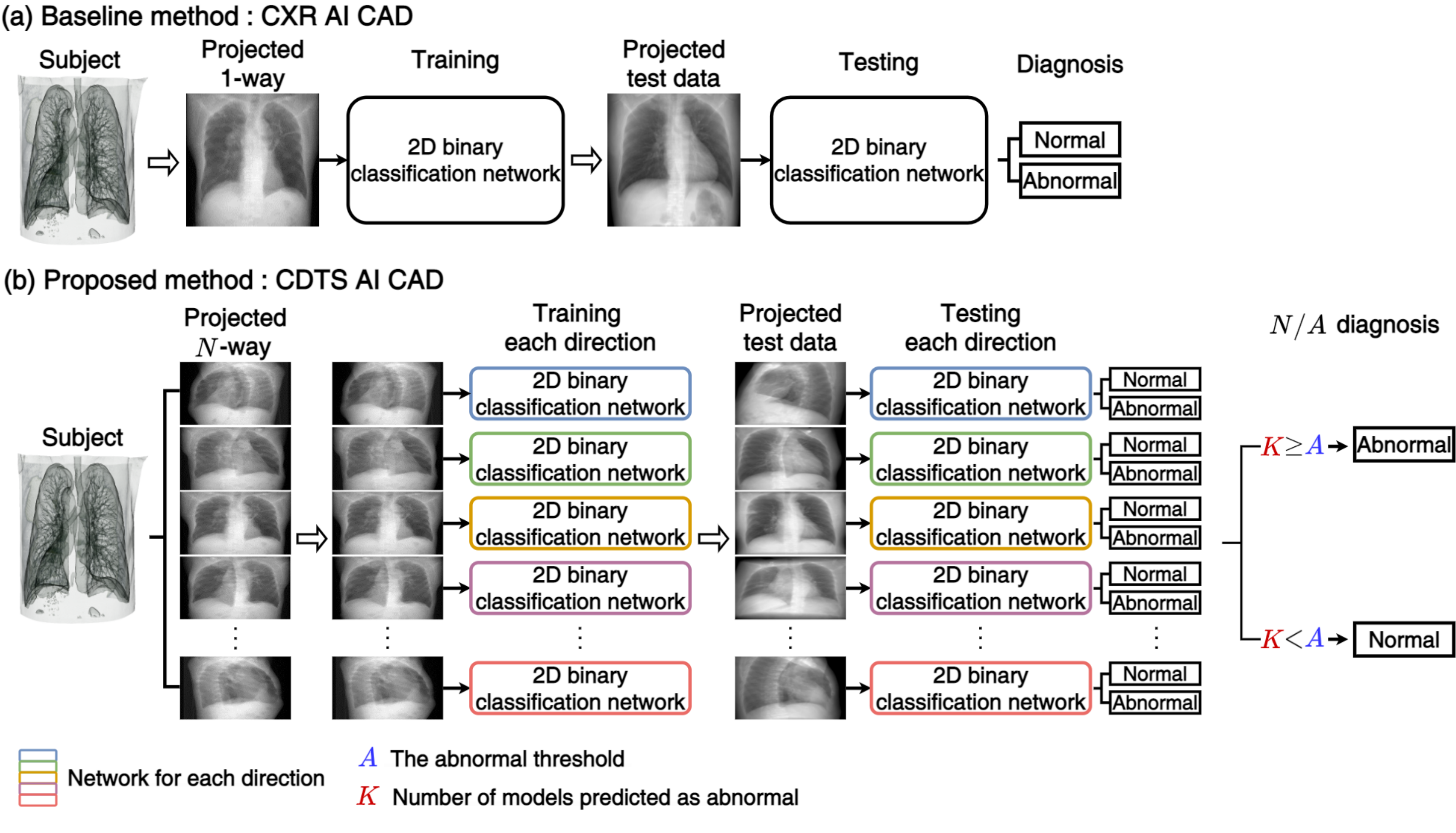}
\caption{Overall architecture of the baseline \& proposed network approach. (a) Baseline method (CXR AI CAD), (b) proposed method (CDTS AI CAD).}
\label{fig:diagnosis_overall}
\end{figure}

\section{Method}
\subsection{Overview of proposed CDTS AI CAD}

\,\,\,\,\,\,\,\, The proposed CDTS AI CAD system is presented in comparison with the baseline CXR AI CAD system in Figure \ref{fig:diagnosis_overall}. Figure \ref{fig:diagnosis_overall}(a) shows the baseline CXR AI CAD technology, which performs diagnosis by receiving a single projection image taken from the front of the target patient as input. This study assumes the diagnosis to be a binary case of normal and abnormal classifications; however, this is sufficiently expanded in subsequent studies. The proposed CDTS AI CAD technology introduced in Figure \ref{fig:diagnosis_overall}(b) receives multiple projection images of CDTS as inputs and performs the same diagnosis as the baseline. An individual network receives each projection image as input and provides a diagnosis result. Our CDTS AI CAD, the final diagnosis result is obtained by synthesizing the multiple diagnosis results received from individual networks (that is, via the proposed $N/A$ diagnosis method, which determines positive if the number $K$ of positives is greater than 
$A$ among $N$ total diagnostic results). In Figure \ref{fig:diagnosis_overall}(b), the number $N$ of multi-projection images is expressed as five, but this number can be set arbitrarily.

The proposed CDTS AI CAD system consists of $N$ networks (i.e., the same number of networks as the number of projected images), and the baseline CXR AI CAD system consists of one network. As this study does not propose a network structure but a learning and inference method, we adopted the existing network WideResNet \cite{zagoruyko2016wide} as the network backbone for the base and proposed models. WideResNet has a wider network width than ResNet \cite{he2016deep}; therefore, we used it because it is more suitable for processing high-resolution images, such as projection images for the lung \cite{wu2019wider}.

\begin{figure}[hbt!]
\centering
\includegraphics[width=0.8\linewidth,height=7cm]{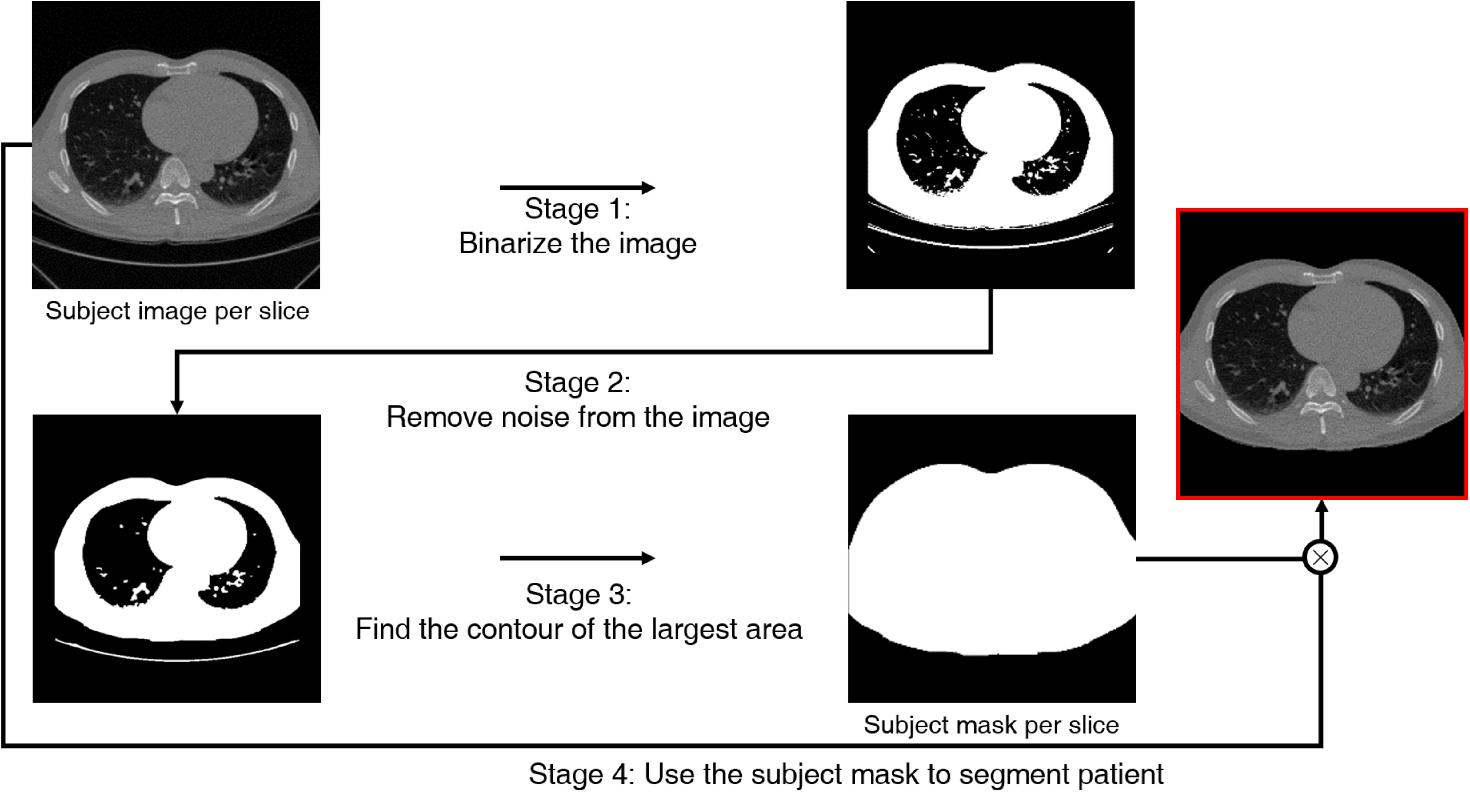} 
\caption{Proposed multi-stage technique to remove scanning bed and extract human volume}
\label{fig:Erase_bed}
\end{figure}

\subsection{Data preparation and processing}
\,\,\,\,\,\,\,\, For a fair comparison between CDTS AI CAD and CXR AI CAD, both CDTS and CXR projection images captured in the same environment for a diseased patient are required. Because of ethical issues this study could only be conducted retrospectively and CXR and DTS projection images captured in the same environment could not be retrospectively and realistically obtained. To obtain CXR and DTS projection images taken in the same environment while solving this problem, we extracted the 3D lung volume from the patient's 3D stack of CT slices and used it as the actual patient's projected target volume. The virtual projection of this volume allows us to capture DTS and CXR projection images in the same environment. \textcolor{black}{Sections 2.2.1 and  2.2.2} show how to prepare the patient's 3D CT stack and extract this 3D lung volume from that stack.
\begin{table}[hbt!]
	\vskip -3pt 
	\caption{\footnotesize {SMC CT dataset used in this study: (Top table) Number of 3D stacks for each lung disease type, (Bottom table) information for each 3D stack  
	}} 
	\footnotesize
	\centering
	{
	\subtable{
		\resizebox{0.6\linewidth}{!}{
			\begin{tabular}{ccccc}
				\toprule
			    Dataset & Label &\:Total\:& \:Segmentation\: & \:Classification\: \\
				\midrule
			
			    \multirow{3}{*}{SMC}
			    &Healthy  &1000& {500} &500\\
				&Pneumonia &242& {-} & 242 \\
				&Tuberculosis  &206&{-} &	206  \\
			
				\bottomrule
			\end{tabular}
			
		} 
      \label{tab:datasets3}
		}
		\vskip -5pt
	\subtable{	
		\resizebox{0.6\linewidth}{!}{
			\begin{tabular}{cc}
				\toprule
			    \small{Parameter} & \small{Value} \\
				\midrule
			   \small{Tube current time product}&  \small{16--139 mAs}\\
			   \small{Tube voltage}&  \small{120 kVp}\\
			   \small{View}& \small{Axial plane}\\
			   \small{Slice size} & \small{512 $\times$ 512 pixels}\\
			   \small{Number of slices}&  \small{135\:--\:390}\\
			   \small{Pixel spacing}& \:\: \small{0.484 $\times$ 0.484${mm}^{2}$\:--\:0.887 $\times$ 0.887${mm}^{2}$}\:\:\\
			   \small{Slice thickness}& \small{0.625, 1, 1.25, 5}\\
				\bottomrule
			\end{tabular}

		} 
	\label{tab:datasets4}
	}
	}
	
	\label{tab:datasets_info}
	\vspace{-0.2cm}
\end{table}

\begin{figure}[hbt!]
\centering
\includegraphics[width=\linewidth]{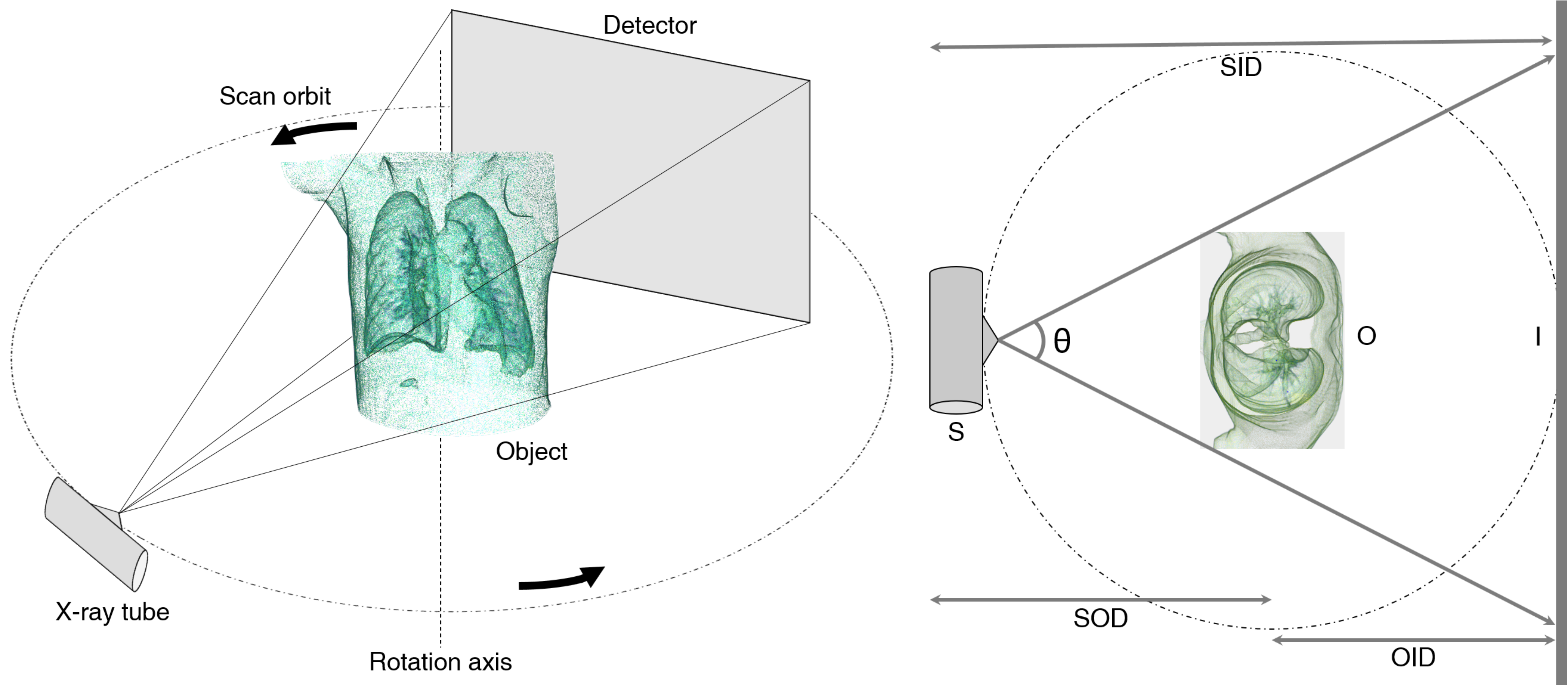}
\caption{Acquisition of multiple DTS projection images through cone-beam projection from the patient's 3D volume. S, O and I indicate the X-ray source, object (subject), and image (detector), respectively, and $\theta$ indicates the fan angle of X-ray beam.}
\label{fig:projection}
\end{figure}

\subsubsection{Data preparation}
\,\,\,\,\,\,\,\, This study was internally verified as a retrospective study, according to the principles of the Declaration of Helsinki and current scientific guidelines. We collected a total of {1448} 3D stacks of DICOM low-dose CT slices from the Samsung Medical Center (SMC), Seoul, Korea, as shown in Table \ref{tab:datasets_info}. The study protocol was approved by SMC's institutional review board (IRB). The {dataset} consists of 3D stacks of CT slices for 1448 patients, that is, 1000 healthy individuals, 242 pneumonia patients, and 206 tuberculosis patients. All data were split to evaluate the proposed AI CAD model and lung region segmentation AI model used in this CAD. We selected 500 cases with normal data to validate the segmentation model and performed five-fold cross-validation on the selected data. To validate AI CAD, we used the rest of the dataset and adopted three-fold cross-validation rather than five-fold to increase the test data more sufficiently.

\subsubsection{Data processing: Acquisition of virtual projection data from CT}
\,\,\,\,\,\,\,\, This section describes the acquisition of virtual projection images from the CT data. This is mainly divided into the process of leaving only the patient's volume from the CT data (i.e., removing other devices such as the bed), and the process of acquiring the projection image from the remaining 3D volume of the patient (i.e., cone-beam projection \cite{scarfe2008cone}). Both the processes are described in the following subsections. 

\subsubsection{Acquisition of patient's 3D volume from CT}
\,\,\,\,\,\,\,\, In actual DTS or CXR imaging, only the patient's {3D} volume is projected. The CT data contained this volume but also included external structures such as beds \cite{kim2008fully,mihaylov2008modeling,zhu2012automatic}, so we removed them to extract only the patient's 3D volume. As illustrated in Figure \ref{fig:Erase_bed}, we propose the following four-stage process of finding the outline of a target person and extracting only the inner area of the person (i.e., removing the CT scanning bed): (Stage 1) We binarized each axial CT slice with a certain threshold. If the pixel value was greater than or equal to the threshold, the pixel was set to 1; otherwise, it was set to 0. For the CT scanning bed to be displayed in the binarized image, we applied a threshold of -500, which is the median of hounsfield units for air and water \cite{pauwels2015cbct}, (Stage 2) In that binarized image, there were cases where the contour of the patient was incorrectly found because of lesions or noise inside the lungs. Therefore, we applied median blur \cite{806630}, and erosion/dilation \cite{1114852} to the binary image to prevent these lesions and noise from interfering with the patient's contour finding process (Stage 3) After founding all contours in the pre-processed/binarized image, we selected the contour of the most significant area and set it as the subject mask. (Stage 4) By multiplying this mask with the actual CT slice, we can leave only the patient's volume. We repeated stages 1–4 for all slices to obtain the patient's 3D volume. External objects, such as beds, can be successfully removed.  

After obtaining the patient's 3D volume through the aforementioned processes, we used the CT's DICOM header information and resized this volume so that its resized distance between adjacent pixels was displayed with a resolution of $1\times1\times1$ mm$^3$.

\subsubsection{Projection data acquisition by virtual cone-beam projection}
\label{projection}
\begin{figure}[t]
\centering
\includegraphics[width=\linewidth]{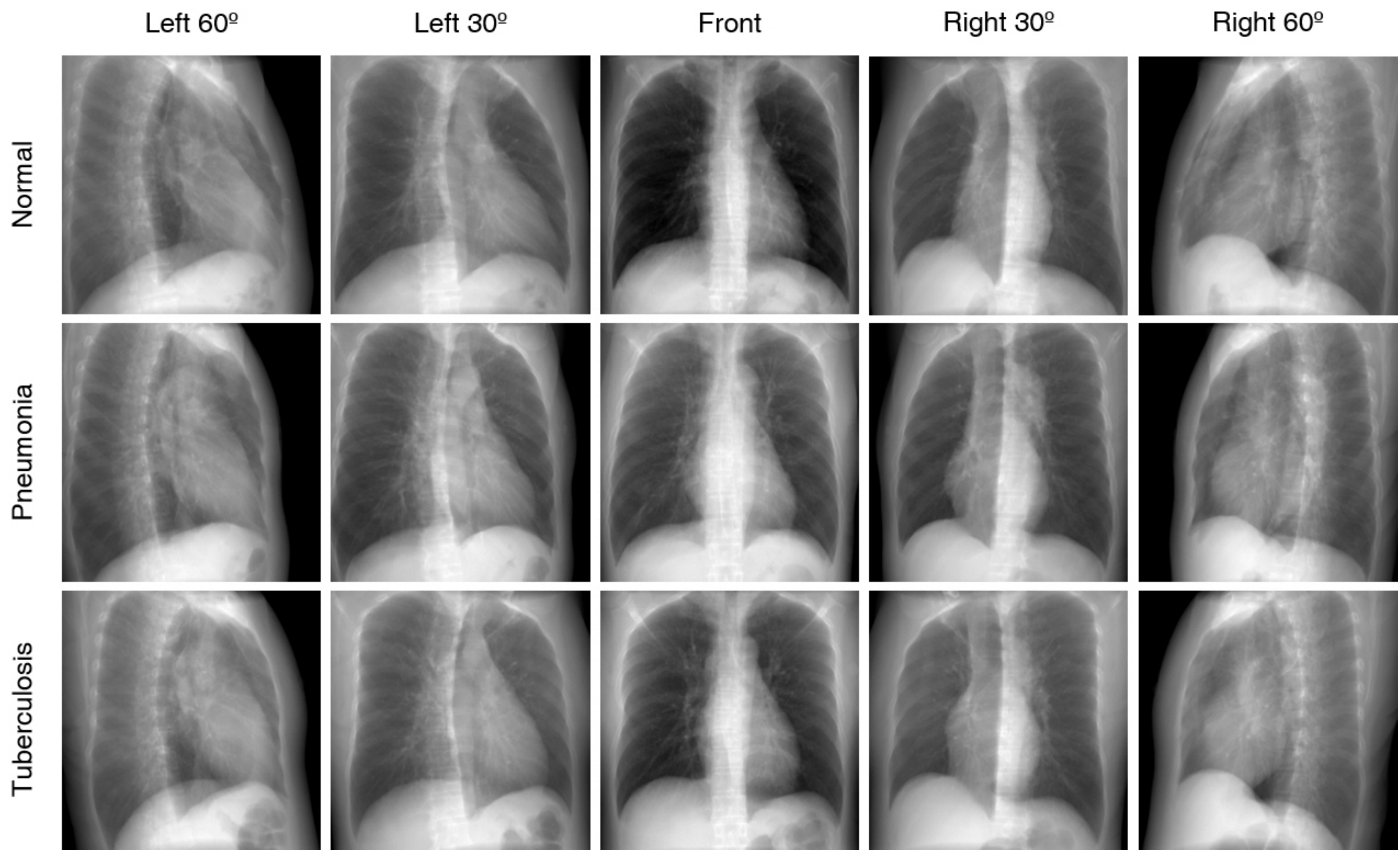}
\caption{Five-way projection images of normal, tuberculosis, and pneumonia patients}
\label{fig:projection_result}
\end{figure}
\,\,\,\,\,\,\,\, After extracting the 3D volume of the target patient, we generated multi-projection images by applying the cone-beam projection technique \cite{scarfe2008cone} to the patient's 3D volume, as shown in Figure \mbox{\ref{fig:projection}}.   
For the corresponding virtual projections, we adopted the digitally reconstructed radiography (DRR) program \cite{milickovic2000ct,staub2013digitally} and acquired five projected images by letting the X-ray camera capture a 3D volume of the patient in the five directions of left 60º, left 30º, front, right 30º, and right 60º degree. For projection geometry, we set the source object distance (SOD, i.e., the distance between the X-ray tube and patient), source image distance (SID, i.e., the distance between the X-ray tube and detector), and object image distance (OID, i.e., the distance between the patient and detector) to 541, 949, and 408 mm, respectively, and we set the size of the detector to 500x500 {mm}. All the geometric information is illustrated in Figure \ref{fig:projection} and an example of the multi-projection image results are presented in Figure \ref{fig:projection_result}.

\subsection{Proposed CDTS AI CAD: Multi-network comprehensive diagnostic method with adjustable disease detection sensitivity }
\begin{figure}[t]
\centering
\includegraphics[width=\linewidth]{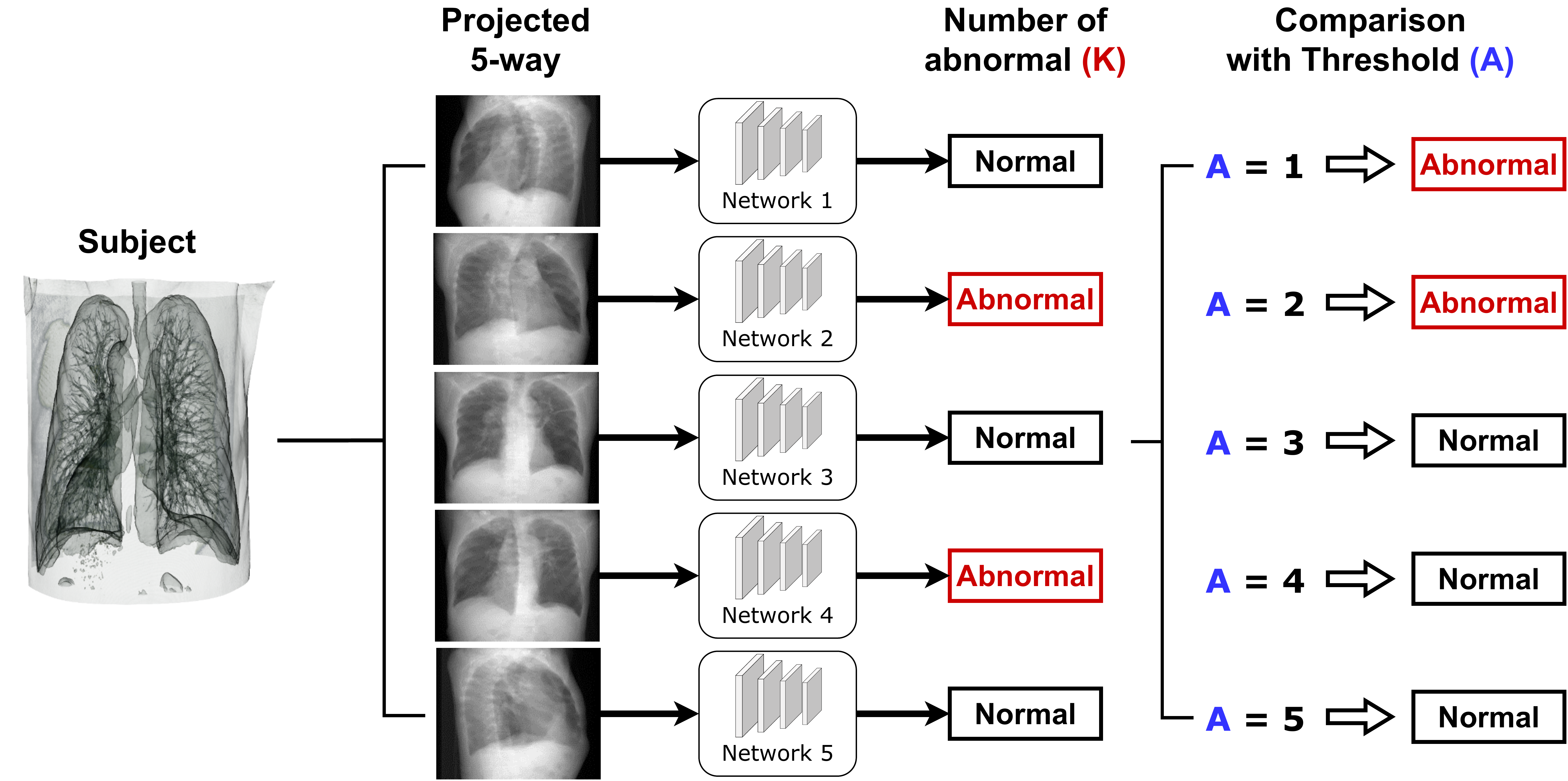}
\caption{Concrete illustration for proposed $N/A$ diagnosis in Figure  \mbox{\ref{fig:diagnosis_overall}}(b): The images projected with $N$ directions are individually taken as inputs for $N$ classification models, each of which diagnoses normal and abnormal. Based on these $N$ diagnosis results, the final result is predicted as normal if there exist over $A$ positive results. In this example, $A$ is set to $2$. }
\label{fig:NA}
\end{figure}

\begin{figure}[t]
\centering
\includegraphics[width=\linewidth]{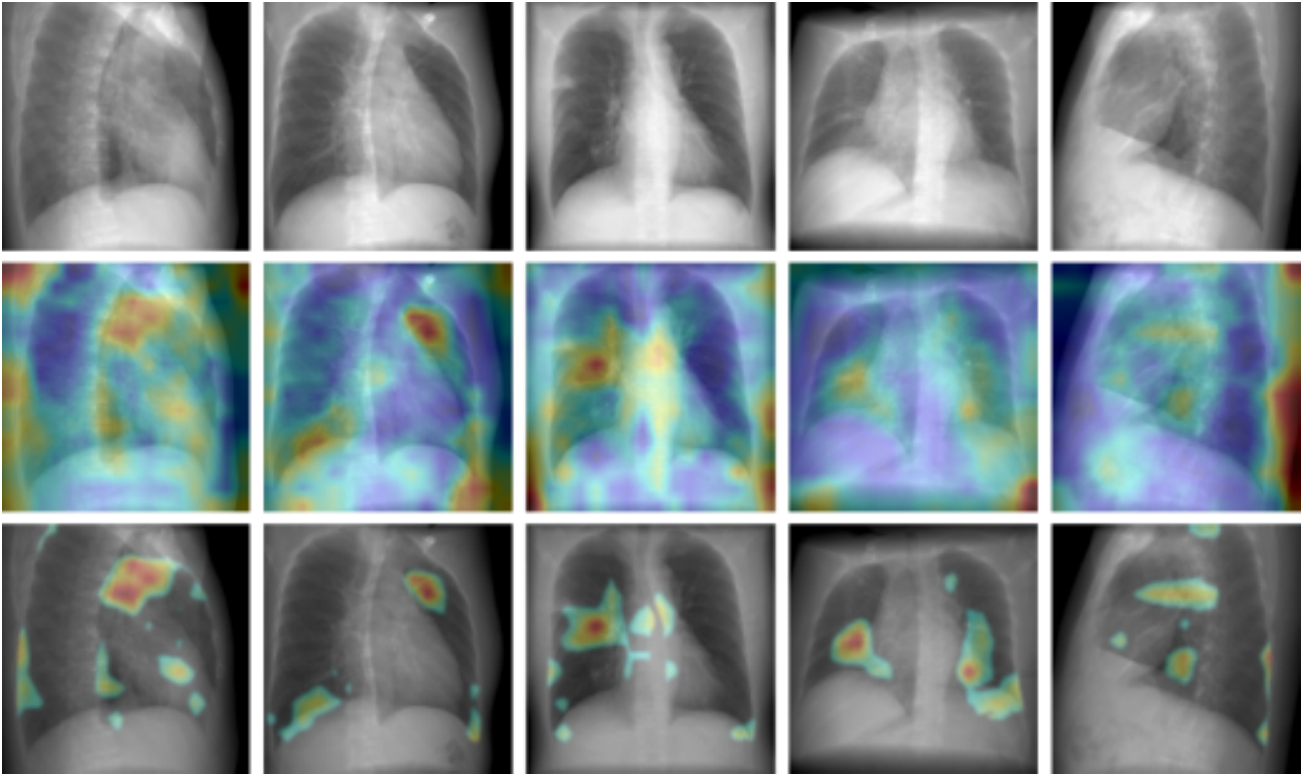}
\caption{The result of lesion (pneumonia) visualization after applying the proposed technique in \eqref{eq_refine_act} (the bottom result). The top result is input projection images and the middle result is a visualization result without applying the proposed technique.}
\label{fig:gradcam_pne}
\end{figure}

\begin{figure}[t]
\centering
\includegraphics[width=\linewidth]{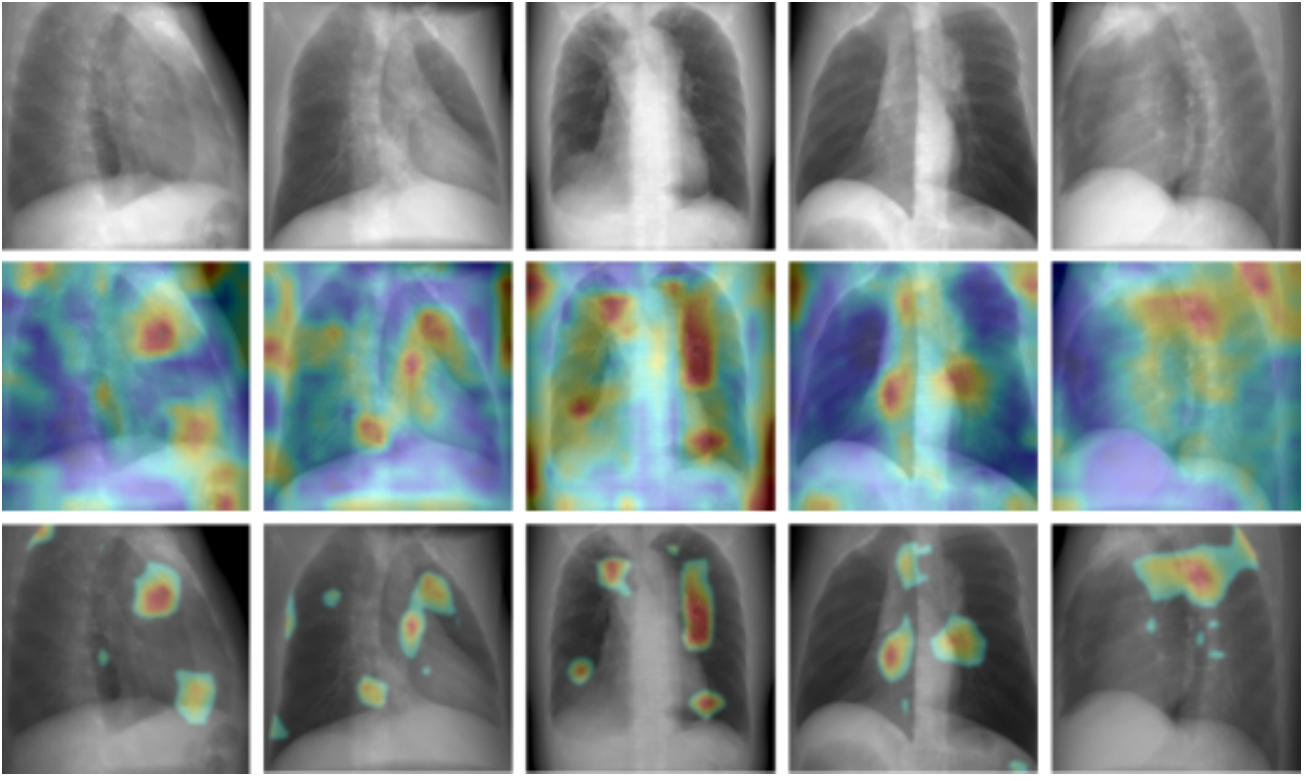}
\caption{The result of lesion (tuberculosis) visualization after applying the proposed technique in \eqref{eq_refine_act} (the bottom result). The top result is input projection images and the middle result is a visualization result without applying the proposed technique.}
\label{fig:gradcam_tb}
\end{figure}

\subsection{Algorithm overview}
\,\,\,\,\,\,\,\, By using $N$ multi-projection images obtained as introduced in \textcolor{black}{Section 2.2}, we developed CDTS-based AI CAD, which consists of $N$ sub-networks taking individual projection images as input and binary classification of whether a patient is abnormal or normal, as shown in Figure \mbox{\ref{fig:NA}}. Specifically, $N$ diagnostic results (\textcolor{blue}{$N$} binary classification results for positive/negative) obtained from independent $N$ subnetworks were integrated through the proposed $N/A$ approach to determine whether the patient was normal or abnormal. Among these multi-projection images, the diagnosis result of a subnetwork using only the frontal image was designated as the baseline CXR-based AI CAD. This study attempted to confirm the superiority of the proposed CDTS AI CAD by comparing its final classification results (obtained from the $N/A$ diagnosis of all subnetworks using all of the multiple projection images) against the baseline CAD classification results.

The proposed $N/A$ method is a method in which the final output result is selected as positive if the number $K$ of all positives among the $N$ total output results is greater than or equal to the predetermined threshold $A$ and negative in other cases. If the network output is given, as shown in Figure \ref{fig:NA}, $K$ becomes $2$. Compared to $K$, when $A$ is $1$ or $2$, the final judgment becomes abnormal, and when $A$ is greater than $2$, the final judgment becomes normal. Our $N/A$ method mainly differs from the existing gold standard ensemble diagnosis method called majority voting. It can adjust the disease sensitivity by adjusting the value of $A$,  further improving the disease sensitivity and achieving higher versatility and practicality.

\subsection{Training details for classification} 
\,\,\,\,\,\,\,\, The proposed CDTS AI CAD and its baseline CXR AI CAD with three-fold cross-validation were trained and tested and performed a classification performance comparison, detailed in \textcolor{black}{Section 3.2.} For the cross-validation, we used the dataset of 3D CT stacks for 948 patients, that is, 500 healthy individuals, 242 patients with pneumonia, and 206 patients with tuberculosis, as shown in Table \ref{tab:datasets_info}. Specifically, model performance was evaluated by three-fold cross-validation for binary classification by selecting one disease among tuberculosis or pneumonia, assuming it was positive, and assuming that the normal group was negative. 

Figure \ref{fig:diagnosis_overall} shows that the proposed CDTS AI CAD has $N$ networks, and each network performs binary classification diagnosis by individually receiving a projection image in each direction as input.We used WideResNet as its backbone \cite{zagoruyko2016wide}. We resized the projection image in each direction to $512 \times 512$ size and applied histogram equalization to use it as an input for each network. For network training, as many studies have been conducted to improve performance, we also adopted the concept of transfer learning \cite{pan2009survey} to use the pre-trained parameters of ImageNet \cite{deng2009imagenet} as the initial parameters of each network. The stochastic gradient descent \cite{ruder2016overview} optimizer with a learning rate of 0.1 was applied as the optimization algorithm, and the multistep learning rate (MultiStepLR) \cite{wu2019demystifying} scheduler was applied to divide the learning rate by 10 for every ten epochs. Cross-entropy \cite{de2005tutorial} was used as the loss function and trained for 50 epochs with a batch size of 64. We also trained the baseline CXR AI CAD using the same setup (i.e., the same setup as the proposed CDTS AI CAD with $N=1$).

\begin{figure}[htb!]
\centering
\includegraphics[scale=0.7]{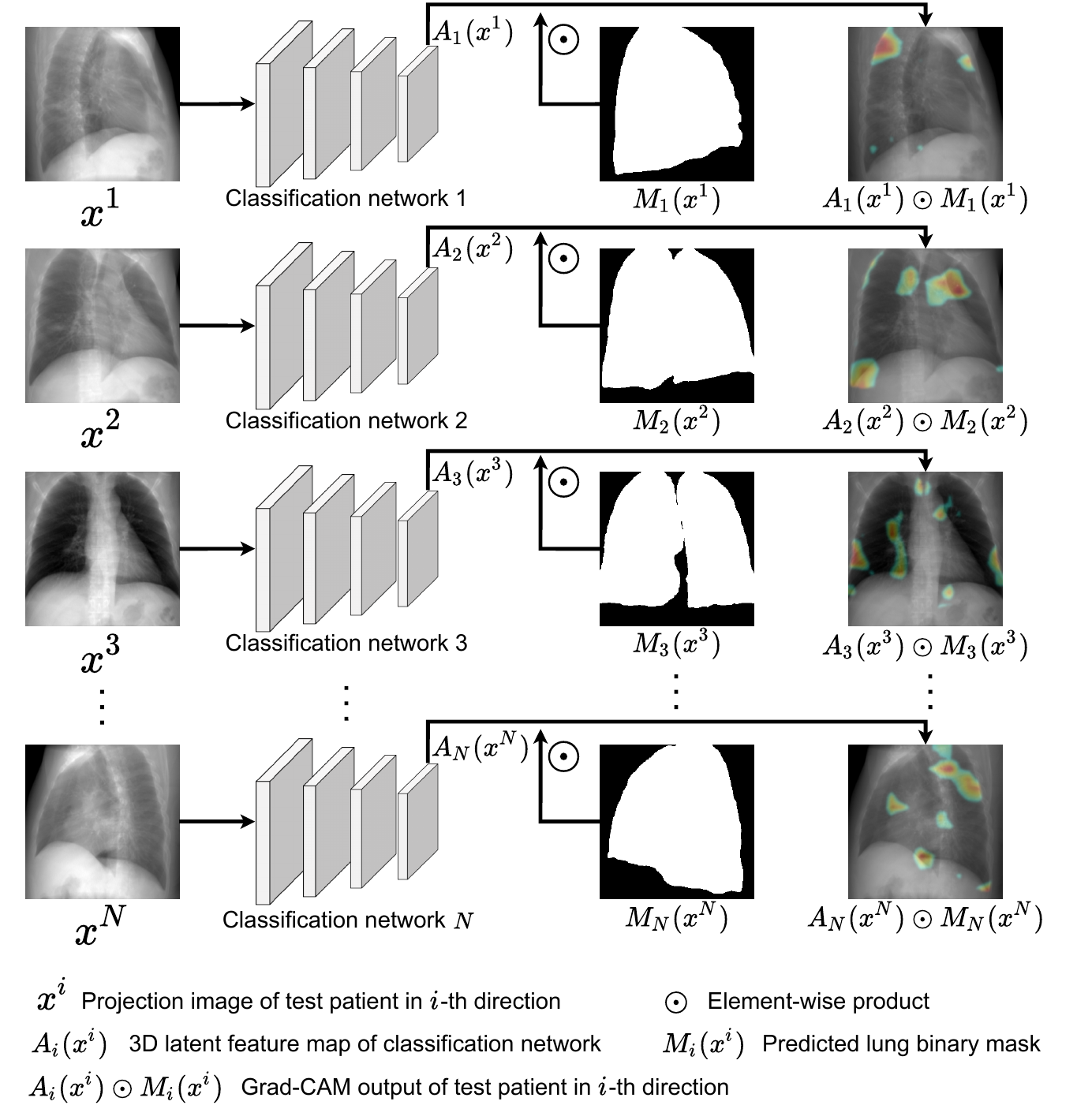}
\caption{Proposed/advanced visualization methods for lesions in each projection image: We infer the lung mask for each projection and reflect it to the network.}
\label{fig:visualization}
\end{figure}

\subsection{Proposed lung lesion visualization for CDTS AI CAD: Enhanced lung lesion visualization through lung area recognition for each projection direction}

\subsection{Algorithm overview}
\,\,\,\,\,\,\,\, This section introduces a technique to visualize lesions so that the CDTS AI CAD network developed, as in \textcolor{black}{Section 2.3}, provides diagnostic results and interprets the results through that visualization. We applied gradient-weighted class activation mapping (Grad-CAM) \cite{selvaraju2017grad} to each projection image-specific model of the proposed CDTS AI CAD to determine which part of the projection image each model observed in the projection image and diagnosed it as a lung lesion/abnormality. The visualization results are shown in the second row of Figures \ref{fig:gradcam_pne} and \ref{fig:gradcam_tb}. Even though a simple application of Grad-CAM to the CDTS AI CAD network, pulmonary lesion visualization results can be obtained for each projection image. These results have the problem of false detection, predicting it as positive even in areas outside the lung.  

\begin{figure}[h]
\centering
\includegraphics[width=\linewidth]{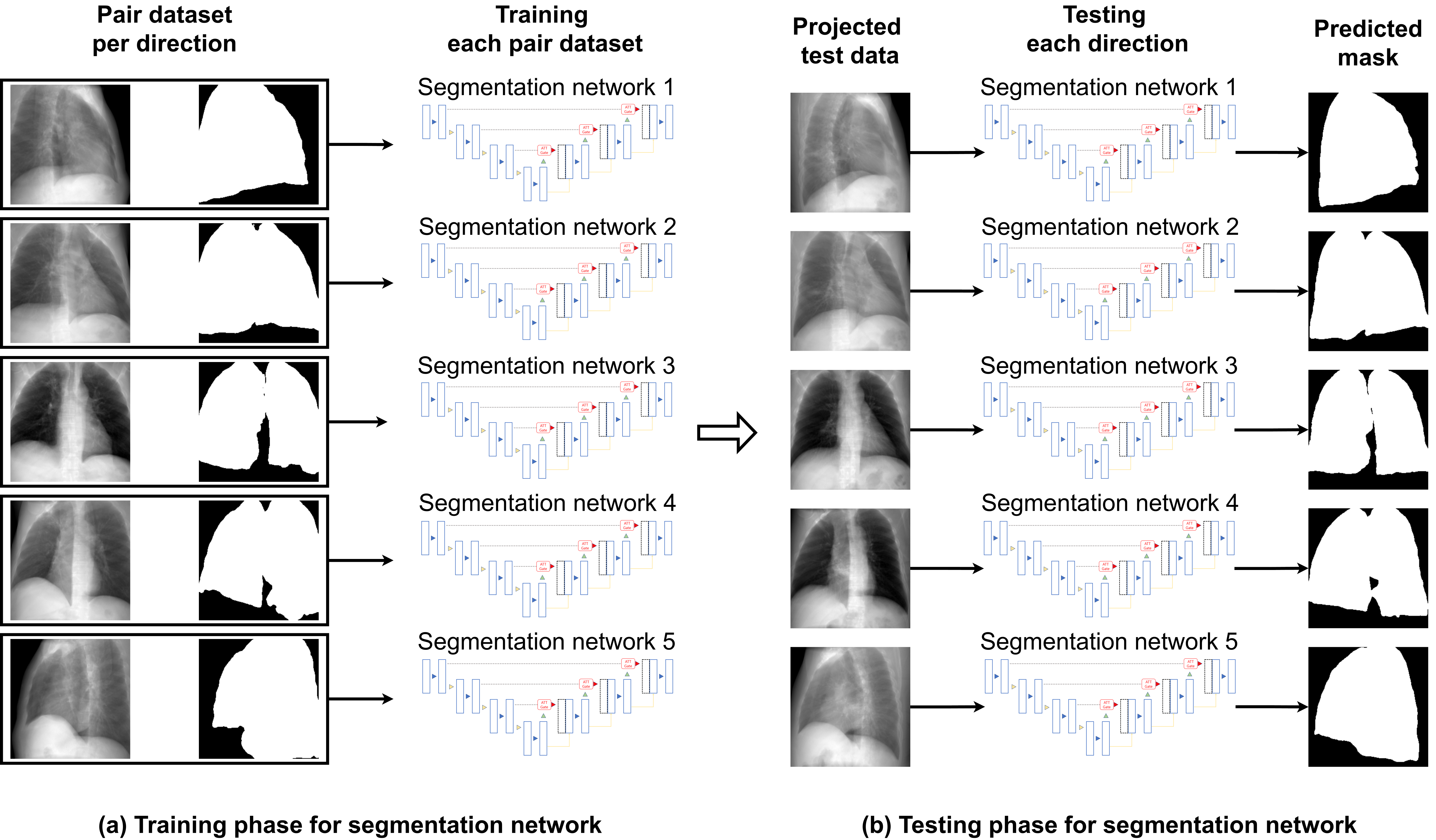}
\caption{The learning (a) and inference (b) process of the segmentation network to automatically generate a mask from the projection image, introduced in Figure \ref{fig:visualization}}
\label{fig:segmentation_mask}
\end{figure}

To address this problem, we forced all the pixel values corresponding to the outer lung area to zero in the latent feature map of the network to which Grad-CAM was applied; thus, the network could only observe and diagnose the inner lung area. The entire process is introduced in Figure \ref{fig:visualization}, where we develop a segmentation network for each projection direction, as shown in Figure \ref{fig:segmentation_mask}, which takes a projection image as the input and outputs a lung area binary mask (matched with each input projection image) to distinguish whether it is the inner lung area. Then, by multiplying the output mask derived from the segmentation network with the feature map and forcing all values for the outer lung area in the feature map to zero, as shown in Figure \ref{fig:visualization}, we can implement such that the CDTS AI CAD network focuses only on the inner lung region, thereby solving the problem of false detection and improving its ability to detect disease visualizations. 

The proposed process for improving the CAM results is presented in Figure \ref{fig:visualization} and is formulated as follows. Suppose that the projection image of the $i$-th direction of a target test patient is denoted by $x^i$ (i.e., $i \in \{1,2,...,N\}$), the 3D latent feature map of the CDTS AI CAD network to which Grad-CAM \cite{selvaraju2017grad} is applied is denoted by $A_{i}(x^i) \in \mathbb{R}^{h \times w \times c}$, and the predicted lung area binary mask, derived as the output of the segmentation network, is denoted by $M_{i}(x^i) \in \{0,1\}^{h \times w \times c}$ (i.e., 0 and 1 for the outer and inner lung areas, respectively). Then, the proposed visualization enhancement technique refines the activation map $A_{i}(x^i)$ of the CDTS AI CAD network, as in \eqref{eq_refine_act}, by using the segmentation network prediction result $M_{i}(x^i)$ for the binary lung mask.

\begin{align}\label{eq_refine_act}
    A_{i}(x^i) \leftarrow A_{i}(x^i) \odot M_{i}(x^i), 
\end{align}
where $\odot$ denotes an element-wise product. 

In the third row of Figures \ref{fig:gradcam_pne} and \ref{fig:gradcam_tb}, we present the visualization results of applying Grad-CAM to feature map $A_{i}(x^i_k)$ updated by \eqref{eq_refine_act}. 
We, therefore, confirmed the validity of this refinement technique in \eqref{eq_refine_act} by showing that the refinement application (the third row) further reduces false detection compared to without using it (the second row). 

\begin{figure}[t]
\centering
\includegraphics[width=0.8\linewidth]{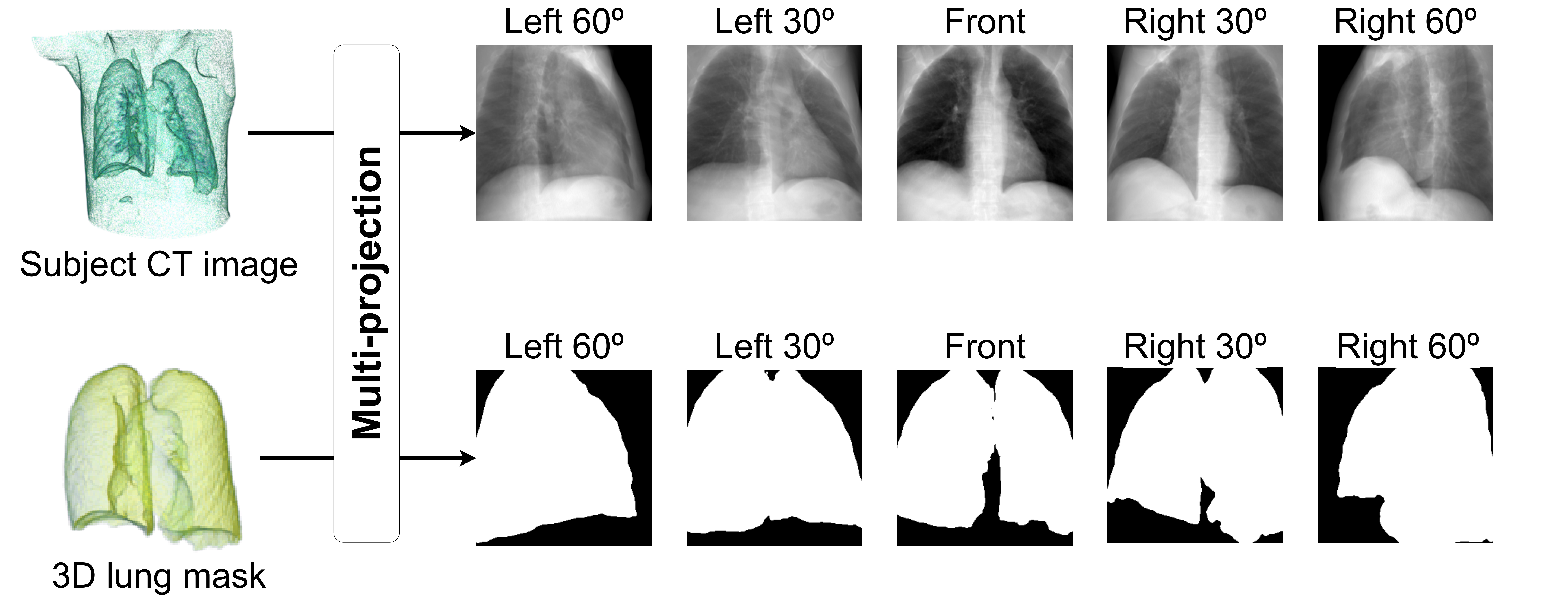}
\caption{Data pair generation technique for training segmentation networks}
\label{fig:pair_dataset}
\end{figure}

\subsection{Training details  for segmentation} 
\,\,\,\,\,\,\,\, This section introduced how the segmentation network used in Figure  \ref{fig:visualization} is trained to provide a lung mask for each projection image. Figure \ref{fig:pair_dataset} shows we extracted the 3D inner region of the lung, binarized the corresponding lung volume, and projected the corresponding binarized volume in the same direction as the actual patient's lung to obtain the corresponding lung mask (i.e., bottom row in Figure \ref{fig:pair_dataset}) for each projected lung image (i.e., top row in Figure \ref{fig:pair_dataset}). To obtain the 3D internal volume of the lung, we used the public library provided by Hofmanninger, J. et al.  \cite{hofmanninger2020automatic}, segmented the 2D lung area for each CT slice, and reconstructed these 2D masks as a 3D mask with the exact dimensions as the patient's body as shown in Figure \ref{fig:pair_dataset}. After generating the training data pair, we set lung X-ray images for each projection direction and their matching lung mask images as the input and output learning labels of each segmentation network so that the network could regenerate the labels as outputs, as illustrated in Figure \ref{fig:segmentation_mask}(a). 

To train $N$ segmentation networks to generate each directional lung mask of CDTS AI CAD, as introduced in Figure \ref{fig:segmentation_mask}(a), we used the following process: (training data pair generation); for 500 patients’ lung 3D CT stacks, as presented in Table \ref{tab:datasets_info}, we obtained $N$ different projection CXR images and their matching binary lung mask pairs, as mentioned above. For $N$ different directions, for simplicity we used the attention-UNet (AUNET) \cite{oktay2018attention}. However, other networks are also available, as this study did not suggest a network. For the input data for each segmentation network, we used the exact pre-processing as used in the classification network,  applied histogram equalization, and resized each image to $256 \times 256$. We used (pixel-level) cross-entropy \cite{de2005tutorial} as the loss function. As the optimization algorithm, we applied the adaptive moment estimation \cite{kingma2014adam} optimizer with a learning rate of 0.0001 and applied the MultiStepLR \cite{wu2019demystifying} scheduler to divide the learning rate by ten at 70, 80, and 90 epochs. We used two batch sizes and trained for 100 epochs. All experiments were conducted on Tesla V100 SXM 32 GB GPUs, and all deep learning models were implemented using Pytorch (v.1.7.0).

Through the learning process in Figure \ref{fig:segmentation_mask}(a), we can obtain ($N$ total) segmentation networks for each projection direction, and we obtained the lung area/mask results corresponding to the X-ray images in each direction, as shown in Figure \ref{fig:segmentation_mask}(b). The segmentation results are evaluated in \textcolor{black}{Section 3.4.}

\section{Results}
\subsection{Performance evaluation metrics}
\,\,\,\,\,\,\,\, We evaluated classification performance according to four statistical analyses: accuracy, precision, sensitivity, and F1 score.
Accuracy indicates the percentage of a total number of test samples in the network identified in the true labels. Precision and sensitivity are denoted as class-wise averages of the proportions detected correctly among all samples detected by the target class and all samples of the target class, respectively, and the F1 score is denoted by the harmonic mean of precision and sensitivity. These four performance measurements were obtained by calculating the true positives (TP), true negatives (TN), false positives (FP), and false negatives (FN) of the confusion matrix constructed by the binary classification model outputs, as follows:

\begin{align}\nonumber  
    Accuracy &= \frac{(TN+TP)}{(TN+TP+FN+FP)},  \\\nonumber 
    Sensitivity &= \frac{(TP)}{(TP+FN)}, \\\nonumber 
    Precision &= \frac{(TP)}{(TP+FP)}, \\\nonumber 
    F1\, score &= 2 \, \frac{Precision \times Sensitivity}{Precision+Sensitivity}.
\end{align}

\begin{table}[ht]
	\vskip -3pt 
	\caption{\footnotesize {Performance comparison between proposed CDTS AI CAD and baseline CXR AI CAD: proposed CDTS AI CAD used multi-projection images ($N/A=5/2$) whereas baseline CXR AI CAD used single-projection image ($N/A=1/1$) }}
	\footnotesize
	\centering
	{
		\resizebox{0.8\linewidth}{!}{
			\begin{tabular}{ccccccc}
            \noalign{\smallskip}\noalign{\smallskip}\hline
				
			Method & Label & Accuracy  &Sensitivity&Specificity &Precision&F1-score \\
			\midrule
			\multirow{2}{*}{CXR AI CAD}
			&Pneumonia  & 
			0.826 $\pm$ 0.010 & 
			0.698 $\pm$ 0.031&
			\textbf{0.888 $\pm$ 0.019} & 
			\textbf{0.752 $\pm$ 0.025} & 
			0.724 $\pm$ 0.016 \\
			&Tuberculosis  & 
			0.874 $\pm$ 0.017 &
			0.728 $\pm$ 0.064 &
	    	0.934 $\pm$ 0.012 &
			0.820 $\pm$ 0.023 &
			0.771 $\pm$ 0.037 \\
			\midrule
			
		    \multirow{2}{*}{CDTS AI CAD}
			&Pneumonia  & 
			\textbf{0.837 $\pm$ 0.027}&
			\textbf{0.785 $\pm$ 0.039}&
			 0.862 $\pm$ 0.028 & 
			0.734 $\pm$ 0.043 & 
			\textbf{0.759 $\pm$ 0.038}  \\
			&Tuberculosis  & 
			\textbf{0.895 $\pm$ 0.028} &
			\textbf{0.782 $\pm$ 0.062}&
			\textbf{0.942 $\pm$ 0.014} & 
			\textbf{0.847 $\pm$ 0.039} & 
			\textbf{0.813 $\pm$ 0.052} \\
			\bottomrule
		\end{tabular}
			
		}
	}
	
	\label{tab:multi_classify}
	\vspace{-0.2cm}
\end{table}
\subsection{Classification performance: Proposed CDTS AI CAD compared to baseline CXR AI CAD}

\,\,\,\,\,\,\,\, We tested the proposed CDTS AI CAD and its baseline CXR AI CAD with three-fold cross-validation and compared the classification performance, as shown in Table \ref{tab:multi_classify}. The proposed CDTS AI CAD uses multi-projection images in five directions as input and derives the final diagnosis result using the $N/A$ method. As the $N/A$ value, $5/2$ was used as a representative, and the results for the other values are introduced in a later section (i.e., Table \ref{tab:Compare_N_a_table}). The baseline CXR AI CAD used only the frontal CXR image among the five multi-projection images, and a value of $1/1$ was used in the $N/A$ method. As shown in Table \ref{tab:multi_classify}, for the binary classification between normal and pneumonia, the proposed CDTS AI CAD improved the baseline CXR AI CAD by 1.1\%, 8.7\%, and F1-score 3.5\%, respectively. For binary classification between normal and tuberculosis, the proposed CDTS AI CAD improved the baseline CXR AI CAD by 2.1 \% accuracy, 5.4 \% sensitivity, 0.8 \% specificity, 2.7\%, and F1-score 4.2\%. These results confirm that the proposed CDTS AI CAD model significantly improves the sensitivity (by 8.7\% and 5.4\% for pneumonia and tuberculosis detection, respectively) compared to the CXR-based model, and the accuracy was maintained.
 
The proposed method using all multi-directional information showed higher performance in diagnosing diseases than the baseline model that only observed the front. The use of multi-directional CDTS projection of multiple images makes it possible to utilize various types of information on lung disease. The proposed AI method effectively classifies diseases compared to the CXR-based AI method that utilizes only a single frontal image, validating the CDTS-based AI CAD.

\begin{figure}[h]
	\centering
	\subfigure[Tuberculosis]{\includegraphics[width=0.45\linewidth, height=5.2cm ]{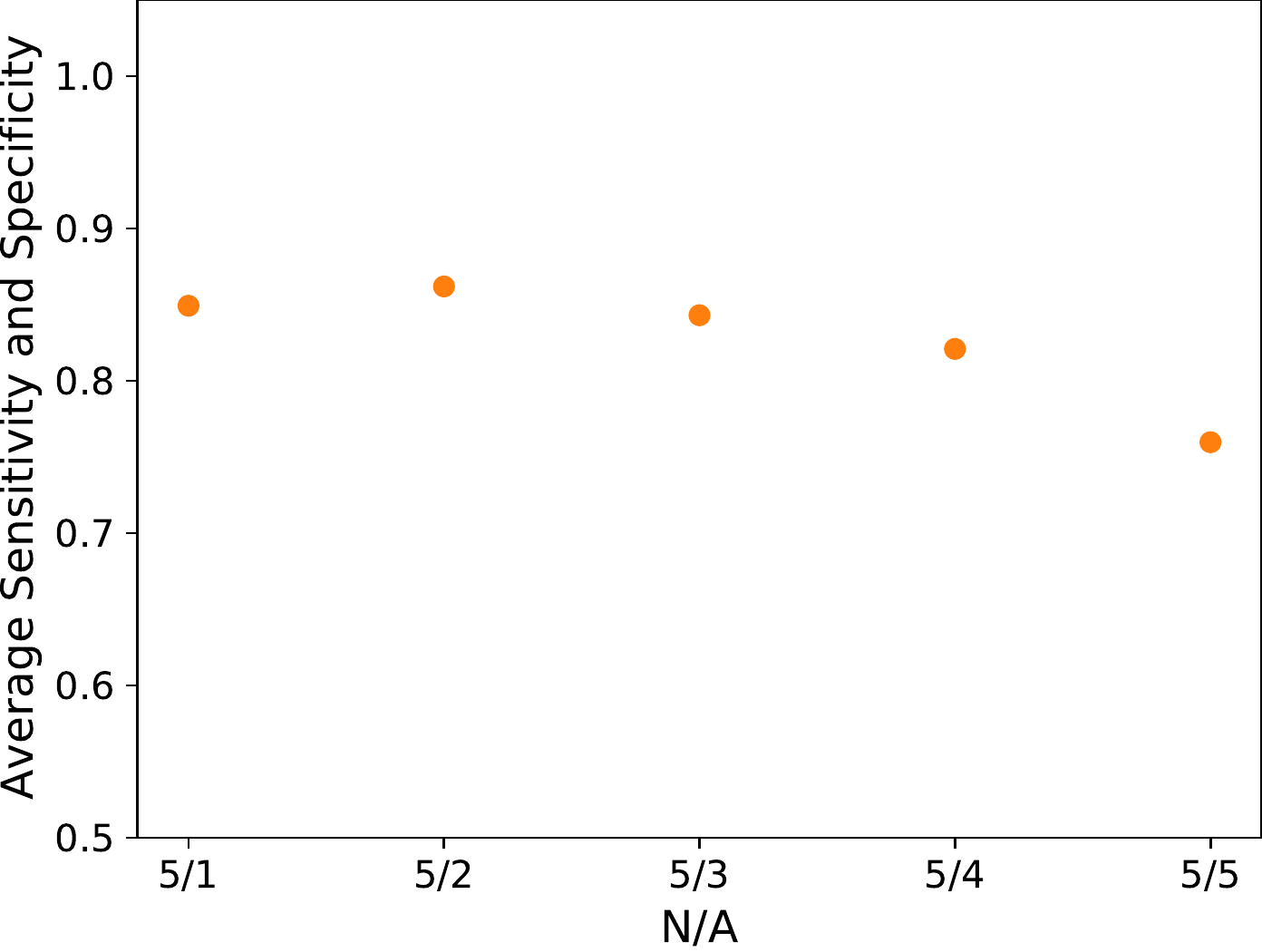}}
	\subfigure[Pneumonia]{\includegraphics[width=0.45\linewidth, height=5.2cm ]{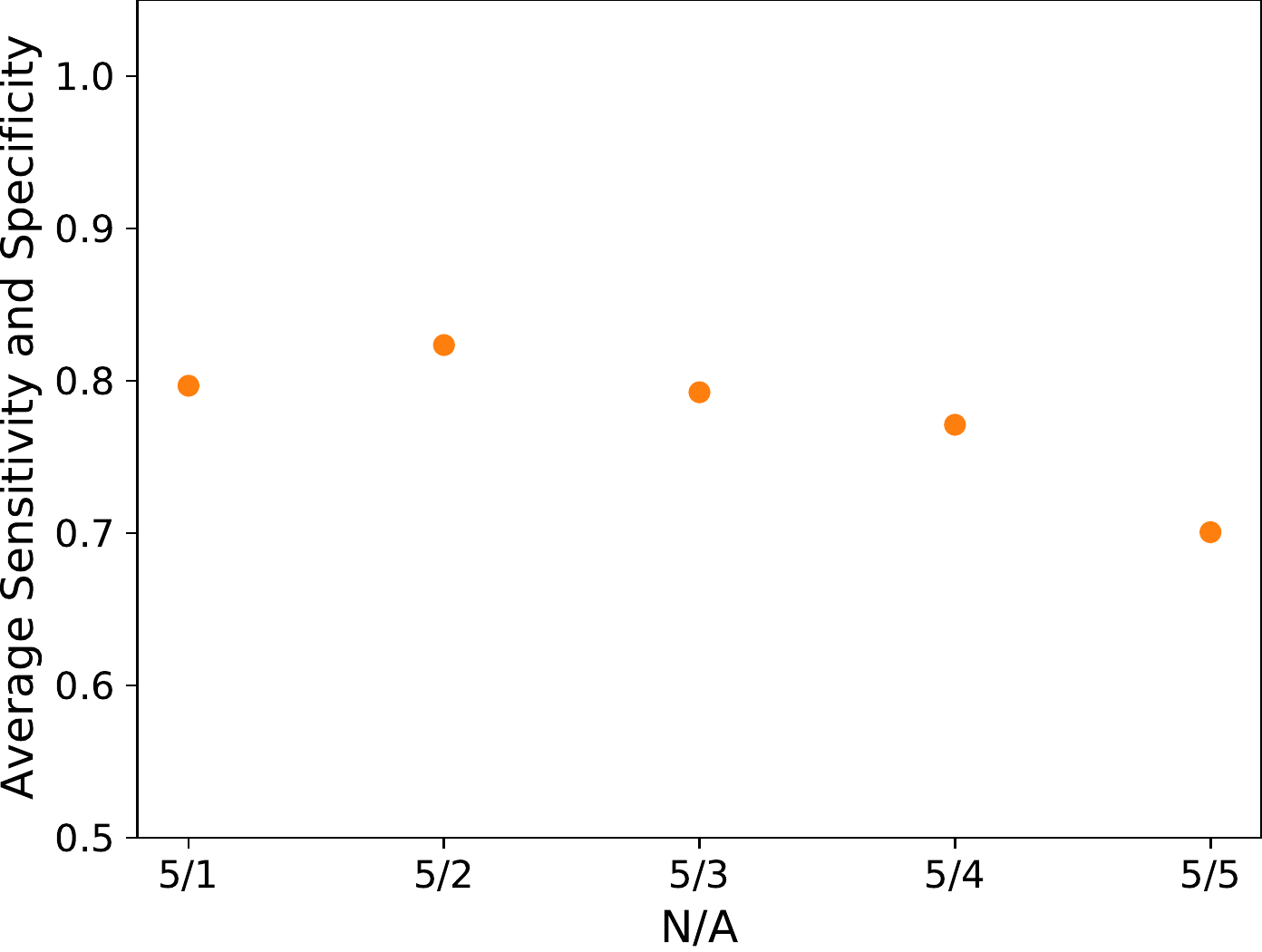}}
	
	\caption{\footnotesize Average values of sensitivity and specificity produced by proposed $N/A$ diagnosis method. For both (a) tuberculosis and (b) pneumonia detections, the average values of sensitivity and specificity were the highest in the $5/2$ diagnostic method.}
	\label{fig:Compare_N_a_fig}
\end{figure}
\begin{table}[ht]
	\vskip -3pt 
	\caption{\footnotesize {The statistical results of classification performance for CDTS-based AI when using $N/A$ method with $N \in \{3, 5\}$ and $A \in \{1,2,...,N\}$. The mean and standard deviation are given by the three-fold cross validation.}}
	\footnotesize
	\centering
	{
		\resizebox{0.9\linewidth}{!}{
	    	\begin{tabular}{ccccccc}
	    	\noalign{\smallskip}\noalign{\smallskip}\hline
			
			    $N{/}A$ & Label & Accuracy & Sensitivity &Specificity& Precision& F1-score \\
				
				\midrule
				\multirow{2}{*}{$3/1$}
        		&Pneumonia  & 
        		0.799 $\pm$ 0.020 & 
        		0.810 $\pm$ 0.027 &
        		0.794 $\pm$ 0.037 & 
        		0.657 $\pm$ 0.037 & 
        		0.725 $\pm$ 0.020 \\
        		&Tuberculosis  & 
        		0.865 $\pm$ 0.015 &
        		0.830 $\pm$ 0.058 &
            	0.880 $\pm$ 0.011 &
        		0.740 $\pm$ 0.015 &
        		0.782 $\pm$ 0.028 \\
        		\midrule
        		\multirow{2}{*}{$3/2$}
        		&Pneumonia  & 
        		0.838 $\pm$ 0.078 & 
        		0.653 $\pm$ 0.057 &
        		0.928 $\pm$ 0.016 & 
        		0.816 $\pm$ 0.019 & 
        		0.724 $\pm$ 0.027 \\
        		&Tuberculosis  & 
        		0.904 $\pm$ 0.029 &
        		0.724 $\pm$ 0.050 &
            	0.978 $\pm$ 0.019 &
        		0.931 $\pm$ 0.060 &
        		0.814 $\pm$ 0.055 \\
        		\midrule
        		\multirow{2}{*}{$3/3$}
        		&Pneumonia  & 
        		0.810 $\pm$ 0.012 & 
        		0.467 $\pm$ 0.052 &
        		0.976 $\pm$ 0.016 & 
        		0.909 $\pm$ 0.048 & 
        		0.615 $\pm$ 0.040 \\
        		&Tuberculosis  & 
        		0.873 $\pm$ 0.007 &
        		0.568 $\pm$ 0.015 &
            	0.998 $\pm$ 0.003 &
        		0.991 $\pm$ 0.015 &
        		0.722 $\pm$ 0.016 \\
        		\midrule
			    \multirow{2}{*}{$5/1$}
				&Pneumonia  & 
				0.774 $\pm$ 0.018 & 
				0.864 $\pm$ 0.026 & 
				0.730 $\pm$ 0.017 & 
				0.608 $\pm$ 0.021 & 
				0.713 $\pm$ 0.023 \\
				&Tuberculosis  & 
				0.847 $\pm$ 0.015 &
				0.854 $\pm$ 0.037 &
				0.844 $\pm$ 0.022 &
				0.694 $\pm$ 0.027 &
				0.765 $\pm$ 0.020 \\
			
				\midrule
				
			    \multirow{2}{*}{$5/2$}
				&Pneumonia  & 
				\textbf{0.837 $\pm$ 0.027} &
				\textbf{0.785 $\pm$ 0.039} &
				\textbf{0.862 $\pm$ 0.028} & 
				\textbf{0.734 $\pm$ 0.043} & 
				\textbf{0.759 $\pm$ 0.038}  \\
				&Tuberculosis  & 
				\textbf{0.895 $\pm$ 0.028} &
				\textbf{0.782 $\pm$ 0.062} & 
				\textbf{0.942 $\pm$ 0.014} & 
				\textbf{0.847 $\pm$ 0.039} & 
				\textbf{0.813 $\pm$ 0.052}  \\
			    
			   	\midrule
	
			    \multirow{2}{*}{$5/3$}
				&Pneumonia  & 
				0.840 $\pm$ 0.002 & 
				0.657 $\pm$ 0.031 & 
				0.928 $\pm$ 0.016 & 
				0.817 $\pm$ 0.025 & 
				0.727 $\pm$ 0.009 \\
				&Tuberculosis  & 
				0.901 $\pm$ 0.025 &
				0.704 $\pm$ 0.042 &
				0.982 $\pm$ 0.018 &
				0.942 $\pm$ 0.058 &
				0.806 $\pm$ 0.049 \\
				
				\midrule

			    \multirow{2}{*}{$5/4$}
				&Pneumonia  & 
				0.841 $\pm$ 0.010 & 
				0.570 $\pm$ 0.058 & 
				0.972 $\pm$ 0.018 & 
				0.912 $\pm$ 0.046 & 
				0.699 $\pm$ 0.033 \\
				&Tuberculosis  & 
				0.894 $\pm$ 0.022 &
				0.646 $\pm$ 0.059 &
				0.996 $\pm$ 0.007 &
				0.984 $\pm$ 0.027 &
				0.779 $\pm$ 0.052 \\
				
				\midrule

			    \multirow{2}{*}{$5/5$}
				&Pneumonia  & 
				0.802 $\pm$ 0.016 & 
				0.409 $\pm$ 0.056 & 
				0.992 $\pm$ 0.009 & 
				0.965 $\pm$ 0.040 & 
				0.572 $\pm$ 0.053 \\
				&Tuberculosis  & 
				0.860 $\pm$ 0.001 & 
				0.520 $\pm$ 0.004 & 
				1.000 $\pm$ 0.000 & 
				1.000 $\pm$ 0.000 & 
				0.684 $\pm$ 0.004 \\
				
				\bottomrule
			\end{tabular}

		}
	}
	\label{tab:Compare_N_a_table}
	\vspace{-0.2cm}
\end{table}
\subsection{Ablation study: Performance comparison of proposed CDTS AI CAD with different numbers of $N$ and $A$}

\,\,\,\,\,\,\,\, The $N/A$ diagnostic method proposed in our CDTS AI CAD is a method of diagnosing as positive (i.e., patient with the disease) if the number of positive predictions among the total $N$ model diagnostic results, which processes $N$ multi-projection images individually, is above the threshold $A$. To validate this $N/A$ method, we conducted a performance comparison of different $N$ and $A$ values, and the results are listed in Table \ref{tab:Compare_N_a_table}. This table presents the comparison results according to different $A$ values for $N=3$ (i.e., projections in the left 60º, front, right 60º) and $N=5$ (i.e., left 60º, left 30º, front, right 30º, right 60º). If we compare the results for two different $N$ values (with the same $A=1$), we can confirm that the sensitivity is higher when $N=5$ than $N=3$. This result indicates that diseases not found in the three views ($N=3$) are more likely to be found from other additional angles ($N=5$), supporting the idea that CDTS (multi-projections) can be superior to CXR (single-projection)-based AI technology. To compare the diagnostic performance according to $A$, we fixed $N$ at $5$ and compared $A$ from $1$ to $5$. Table \ref{tab:Compare_N_a_table} shows the F1 score for each detection of pneumonia and tuberculosis was the highest at $A=2$. We also present the arithmetic mean value of sensitivity and specificity in Figure \ref{fig:Compare_N_a_fig}, which is also the highest at $A=2$, similar to the F1 score, supporting the overall excellence when $A=2$. We developed a CDTS AI CAD model by selecting the value of $N/A$ as $5/2$.

Note that $N/A$ value of $5/2$ has higher accuracy than the other $A$ however, $5/1$ has a higher sensitivity than $5/2$ as given in Table \ref{tab:Compare_N_a_table}. This implies that selecting a lower $A$ is more effective if one needs to increase disease sensitivity. 

\begin{table}[ht]
\caption{\footnotesize {Segmentation performance in each direction (e.g., $N=5$) in terms of JS, IoU and DICE of reference study and our study. The mean and standard deviation are given by the five-fold cross validation.}}
\centering
{
	\resizebox{.7\linewidth}{!}{
		\begin{tabular}{cccc}
            \noalign{\smallskip}\noalign{\smallskip}\hline
			
			  {} & JS & IOU & DICE\\
			\midrule
			 Front & 0.953 $\pm$ 0.003 & 0.975 $\pm$ 0.001 &0.953 $\pm$ 0.001 \\
			 Left side30 & 0.965 $\pm$ 0.001 & 0.981 $\pm$ 0.001 &0.991 $\pm$ 0.001 \\
			 Left side60 & 0.969 $\pm$ 0.001 & 0.979 $\pm$ 0.001 &0.989 $\pm$ 0.001 \\
			 Right side30& 0.963 $\pm$ 0.002 & 0.977 $\pm$ 0.001 &0.990 $\pm$ 0.001 \\
			 Right side60& 0.969 $\pm$ 0.002 & 0.978 $\pm$ 0.001 &0.989 $\pm$ 0.001 \\

			\bottomrule
		\end{tabular}
	}
}
\label{tab:segmentation_eval}
\vspace{-0.2cm}
\end{table} 

\begin{figure*}[ht]
	\vskip -6pt
	\centering
	\subfigure[Normal]{\includegraphics[width=0.3\linewidth,height=7cm]{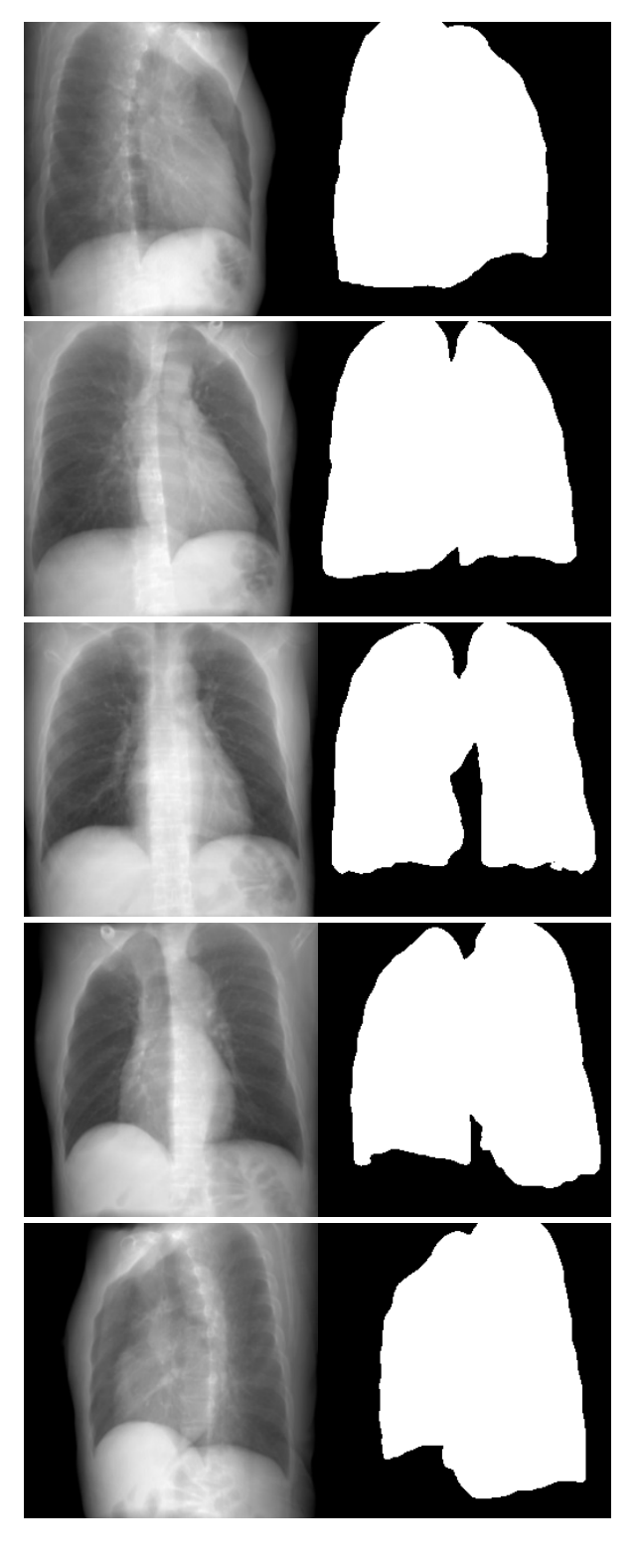}}
	\subfigure[Pneumonia]{\includegraphics[width=0.3\linewidth,height=7cm]{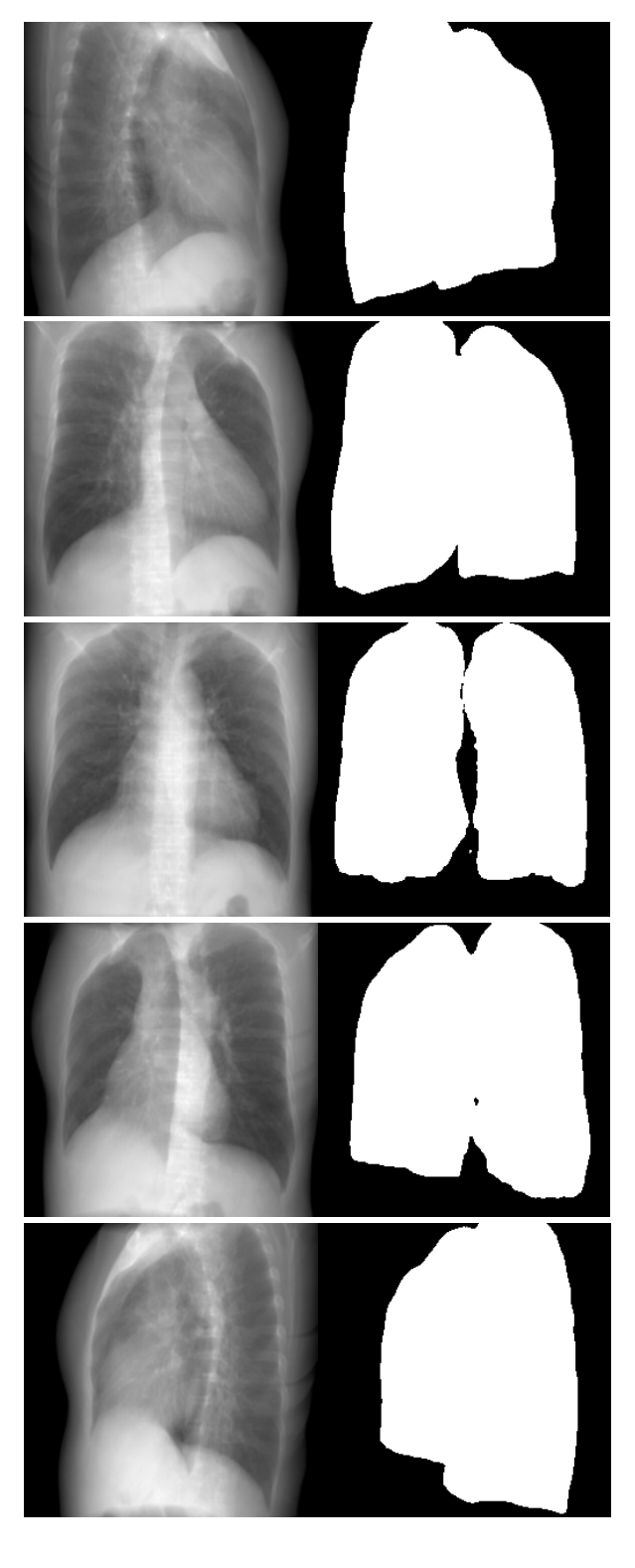}}
	\subfigure[Tuberculosis]{\includegraphics[width=0.3\linewidth,height=7cm ]{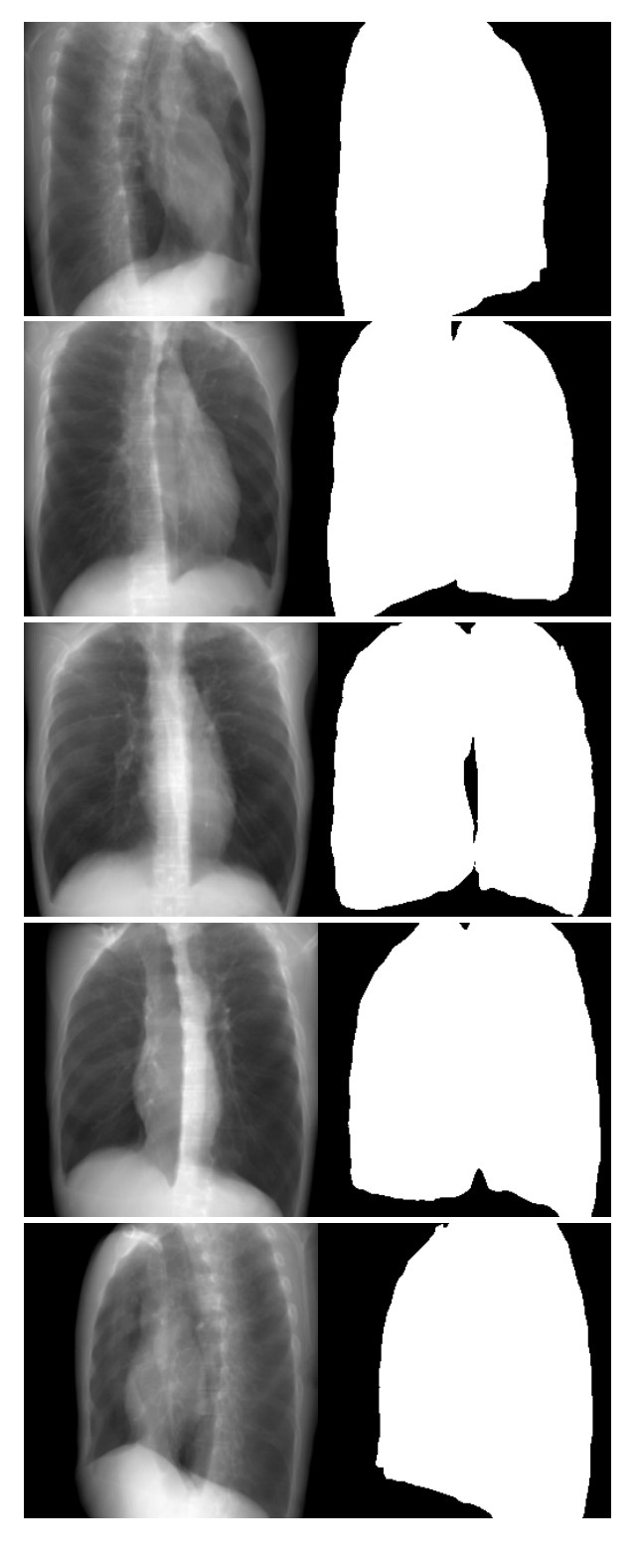}}
	\vskip -6pt
	\caption{\footnotesize  {Example of lung mask predicted by the segmentation network for each projection direction (e.g., $N=5$)}}
	\label{fig:sup_comp2}
\end{figure*}

\subsection{Segmentation performance: Lung mask prediction results internally used in proposed CDTS AI CAD}
\,\,\,\,\,\,\,\, This section evaluates the lung-mask prediction results produced by the segmentation networks (i.e., AUNET). A total $N$ segmentation networks, which take projection images for each direction and predict its matching lung mask, as illustrated in Figure \ref{fig:segmentation_mask}(b), were individually trained and evaluated with five-fold cross-validation. The evaluation was performed using the Jaccard similarity coefficient (JS), intersection over union (IoU), and Dice coefficient (DICE) for the binary lung mask results predicted by each directional/trained model. Table \ref{tab:segmentation_eval} presents the performance results and the example is presented in Figure \ref{fig:sup_comp2}. These results show that all scores in each direction are 0.95 or higher, indicating that our trained segmentation network shows sufficiently good segmentation performance, and these figures also support our observation.

\section{Discussion}
\,\,\,\,\,\,\,\, Recently, many studies \cite{sitaula2021attention,le2021iot,chen2020label,mittal2020detecting,siddiqi2020efficient} have introduced CXR AI CAD technology: Sitaula et al. \cite{sitaula2021attention} advanced  the conventional CXR AI CAD by adopting the attention technology, while Dac-Nhuong Le et al. \cite{le2021iot} introduced a deep support vector machine to detect COVID-19 on CXR,  Chen et al. \cite{chen2020label} addressed multiple disease classification on CXR via graph convolution networks, and Mittal et al. \cite{mittal2020detecting} and Siddiqi et al. \cite{siddiqi2020efficient} also developed new convolutional neural network (CNN)-based models more suitable for CXR diagnosis. However, the existing CXR AI CAD model is not scalable with the CDTS AI CAD model. Our method was designed to allow this extension to perform a fair comparative analysis of the performance of CDTS AI CAD and CXR AI CAD. Our study suggests a methodology to extend the existing CXR model to a CDTS AI CAD model. Thus, any existing CXR model is used as the backbone of the CDTS/CXR AI CAD model. It is possible to further advance the CDTS AI CAD performance by applying the latest CXR AI CAD models that will appear in the future. 

Because of the scalability between CXR and CDTS of the proposed technology, we could compare the performance of CXR and CDTS-based AI CADs, going beyond the previous study that compared the performance between CXR and CDTS only based on the human diagnosis. Sufficient studies show that CDTS is superior to CXR from a clinical perspective (i.e., due to a physician's diagnostic comparison). Lee et al. \cite{lee2013comparison} argued that CDTS (sensitivity 82\%) was more sensitive than CXR (sensitivity 27\%) for asbestos-related pleuropulmonary disease detection in thoracic radiologist reading criteria. Galea et al. \cite{galea2014practical} found that CDTS provides better sensitivity than conventional CXR for lung nodule detection by reducing the anatomical noise or composite artifacts. Quaia et al. \cite{quaia2010value} reported that CDTS improved both diagnostic reliability and accuracy compared with CXR. Kim et al. \cite{kim2016comparison} demonstrated that CDTS
(sensitivity 86\%) showed a better lung nodule detection performance than CXR (sensitivity 61\%). As such, the superiority of CDTS compared to CXR has been sufficiently reported but only based on the human diagnosis. This study compared and proved it based on AI CAD; therefore, it supports the superiority of CDTS and existing clinical results.
 
The study most similar to ours was the one suggested by \textcolor{black}{S. Chauvie et al.} \cite{Chauvie2020ArtificialIA} in that they also developed CDTS AI CAD. They developed AI CAD based on images after 3D reconstruction of the CDTS and confirmed the validity of lung cancer detection. However, their study may result in significant information loss \cite{sidky2009enhanced,velikina2007limited,hu2008image} in the signal processing reconstruction process in terms of using samples after conventional DTS reconstruction. Unlike their research, our study can prevent information loss because it uses a sample before restoration. Our CDTS AI CAD can be directly extended to AI CAD for CXR, thereby enabling comparison with CXR to demonstrate its superiority over CXR for the first time. Alternatively, various studies \cite{samala2016mass,samala2018breast,xiao2021classification,zhang20192d} have been conducted as DTS AI CAD for breasts, not DTS AI CAD for chest, but these are all based on images after 3D restoration. Therefore, their studies are fundamentally different from techniques based on pre-reconstruction images like ours.  

Our study had several limitations. This study considered only values of $3$ and $5$ for the number $N$ of CDTS projected image directions. To prove that this is excellent, we proceeded in the direction of $N=5$, but it may be necessary to further increase $N$ depending on the lesion or data. The CDTS used in our study uses $N$ multi-projection images. For simplicity, we represent $N$ as $5$ to emphasize that CDTS (i.e., $N>1$) can outperform CXR (i.e., $N=1$) -based CAD. However, depending on the lesion or data type, further increases in $ N$ are expected to be necessary to increase $N$ in practice further. In addition, as the projected image from the actual DTS machine was not used, we plan to prospectively collect clinical data as a follow-up study and try further validation using this technology.

\section{Conclusion}
\,\,\,\,\,\,\,\, This study proposes a novel CDTS AI CAD based on multi-projection/pre-reconstruction images. This enabled a fair performance comparison with the existing CXR AI CAD, thereby demonstrating that CDTS AI CAD significantly improved pneumonia and tuberculosis detection sensitivity to CXR AI CAD, which are representative lung diseases. We proposed a methodology to effectively reduce the false positive rate in visualizing the lesion location in each projection image of DTS and confirmed its effectiveness. These results demonstrate that our CDTS AI CAD method can provide doctors with higher diagnostic performance and additional auxiliary lesion visualization data using only a dose similar to that for the existing CXR AI CAD. Therefore, we expect that CDTS AI CAD-based technology will be an effective alternative to CXR-based AI CAD for the initial diagnosis of lung abnormalities. 

\section*{Acknowledgements}
\,\,\,\,\,\,\,\, This work was supported by the National Research Foundation of Korea (NRF) grant funded by the Korean government (MSIT) (2021R1F1A106153511). This work was also supported by the Korea Medical Device Development Fund grant funded by the Korean government (Ministry of Science and ICT, Ministry of Trade, Industry and Energy, Ministry of Health $\&$ Welfare, Ministry of Food and Drug Safety) (202011B08-02, KMDF$\_$PR$\_$20200901$\_$0014-2021-02). This work was also supported by the Technology Innovation Program (20014111) funded by the Ministry of Trade, Industry $\&$ Energy (MOTIE, Korea). This work was also supported by the Future Medicine 20*30 Project of the Samsung Medical Center (SMX1210791).

\clearpage 
\bibliographystyle{unsrtnat}
\bibliography{ref.bib}
\clearpage 

\appendix
	
\end{document}